# The Tunneling Hamiltonian Representation of False Vacuum Decay: IV. Application to formation of Dark Matter in nucleating universe

A.W. Beckwith

Department of Physics and Texas Center for Superconductivity

University of Houston Houston, Texas 77204-5005, USA

**Abstract**

The tunneling Hamiltonian has proven to be a useful method in many-body physics to treat particle tunneling between different states represented as wavefunctions. Here we apply a generalization of the way we formed appropriate wave functionals for CDW to how to present nucleation of an inflationary universe . This allows us to make a generalization of the model used for inflation where the inflaton is trapped in a false vacuum and , after tunneling makes a first order phase transition to initiate nucleation of an inflationary universe , in which tunneling between states which are wavefunctionals of a scalar quantum field $\phi$ are considered. We explain how we can have particle – anti particle pairs as a model of how nucleation occurs and construct a potential which may permit formation of dark matter using Sheherrers k-essence model construction .

Correspondence:

A. W. Beckwith:   **projectbeckwith2@yahoo.com**

PACS numbers   : 03.75.Lm, 11.27.+d, 98.65.Dx, 98.80.Cq, 98.80.-k

# I. Introduction

In our paper, we have managed to show commonality of a specific interpretation of chaotic inflationary cosmology presented by Guth[1] which builds upon a presentation given by Linde[2] with a false vacuum model of the nucleation of the universe which was initially pioneered by Coleman[3] and then greatly refined by Garriga[4] for the problem of nucleation of an electron- positron pair in a de Sitter cosmology , by use of the Bogomil'nyi inequality[5] to shape a wave functional which could be used to be a nucleation rate value for particle creation for a unit length of space time. We set that unit value of spacetime to be the planck length and then suggested how this procedure could tie in with a tunneling Hamiltonian representation of current 'density' [6] from a nucleating universe . This last step is incomplete but with additional developments we think it will show the fundamental physics of creation of our universe from the big bang synthesis of matter as we know it. Furthermore, we should note that this model appears to be congruent with the existence of a region which is the flat slow roll requirement of $\left|\frac{\partial^2 V}{\partial \phi^2}\right| << H^2$ where $H$ is the expansion rate which is a requirement of realistic inflation models [4]. This is necessary and contravenes earlier results by Coleman-de Lucia [3,4] which appears to show this is impossible.

We also , afterwards have formed , using Scherrers recent article [7], a template for evaluating initial conditions which would shed light on if or not this model universe would be radiation dominated in the beginning, or would be more in sync with having dynamics determined by assuming a straight cosmological constant all throughout expansion of this universe. We will evaluate this question in full in a subsequent paper.

## I. Chaotic inflationary scenarios and their tie in with our problem.

Guth [1] as of 2000 wrote two well written articles with regards to the problem of the basic workings of inflationary models, as well as summaries as for why our universe is the product of inflation. The simplest of these models, called the chaotic inflationary model [1] via use of a massive scalar field construction gives an elegant treatment of how we could have an inflation field $\phi$ set at a high value $\phi \equiv \tilde{\phi}_0$ and which then would have an inequality of

$$\tilde{\phi}_0 > \sqrt{\frac{60}{2 \cdot \pi}} M_P \approx 3.1 \cdot M_P \tag{1}$$

This pre supposes a harmonic style potential of the form [2]

$$V \equiv \frac{1}{2} \cdot m^2 \cdot \phi^2 \tag{2}$$

where we have classical and quantum fluctuations approximately giving the same value for a phase value of [1]

$$\phi^* \equiv \left(\frac{3}{16 \cdot \pi}\right)^{1/4} \cdot \frac{M_P^{3/2}}{m^{1/2}} \cdot M_P \rightarrow \left(\frac{3}{16 \cdot \pi}\right)^{1/4} \cdot \frac{M_P^{3/2}}{m^{1/2}} \tag{3}$$

where we have set $M_P$ as the typical plancks mass which we normalized to being unity in this paper for the hybrid false vacuum – inflaton field cosmology example, as well as having set the general evolution of our scalar field as having the form of

$$\phi \equiv \tilde{\phi}_0 - \frac{m}{\sqrt{12 \cdot \pi \cdot G}} \cdot t \tag{4}$$

We in writing this do not have any clearly defined false vacuum for our problem as would be intuitively want to have, and we are going to in the new publication incorporate some of the insights of the chaotic inflation model, including a favored $\phi \equiv \phi^*$ as a reference point to what we will do in our treatment of forming a Gaussian presentation of a wave functional which will incorporate a first order phase transition. In addition, we also will in this treatment use a 'toy model' for a 1 + 1 dimensional presentation of nucleation assuming as we do that after a brief instant of planck time $t_P$ that we have had a 'pop up' of a comparatively electrically neutral **S-S'** formation of matter which will then be assumed to have a separation of $L$ between its constituent parts which will be a first order approximation to the 'radius' of the universe at the start of inflationary expansion. See **appendix entry I** for the Bogomil'nyi inequality contribution to forming this wavefunctional which we used in our derivation.

**I   Description of the potential plus Lagrangian model used in our wavefunctional**

We begin this by considering a Lagrangian 'density' of the form

$$L = \frac{1}{2}\left(\partial_\mu \phi \ \partial^\mu \phi\right) - V(\phi) \tag{6}$$

where we are considering a continuous scalar field and where we are looking at reasonable potentials which would incorporate some of the insights of the chaotic inflation model ( a $\phi^2$ potential dependence ) with false vacuum nucleation. For our potential, we worked with [8,9,10]:

$$V_1(\phi) = M_P^2 \cdot 5989 \cdot (1 - \cos(\phi)) + \frac{m^2}{2} \cdot (\phi - \phi^*)^2 \tag{7}$$

where $M_P \gg m$ as well as an overall potential of the form

$$V(\phi) \equiv \left[initial \ \ energy \ \ density\right] + V_1(\phi) \tag{8}$$

where what we are calling the *initial energy density* is a term from assuming a brane world type of potential usually written as [8,9,10]

$$V(\phi,\tilde{\psi}) = \frac{1}{4}\cdot(\tilde{\psi}^2 - M_P^2)^2 + \frac{1}{2}\cdot\lambda'\cdot\phi^2\cdot\tilde{\psi}^2 + V_1(\phi) \qquad (9)$$

where the 'radial' component $\tilde{\psi}$ is nearly set equal to zero and the scalar potential in our case is changed from a $\phi^2$ potential dependence to one where we incorporate a false vacuum nucleation procedure as given by $V_1(\phi)$. In order to do this we will look at setting values of $\phi \equiv \phi^*$ [1] due to the chaotic inflation model [1,2] and then consider a specific ratio of $M_P$ to mass $m$ to work with. In order to get our model congruent with respect to the Bogomil'nyi inequality results [5,6,9] outlined in appendix I and the first paper of this series, we set $M_P \cong 2.269\cdot m$ so as to have the following still hold ( if we use the 'normalization of setting $M_P = 1$ )

$$\phi_F \cong .5472 \qquad (10)$$

and

$$\phi_T \cong 5.457 \qquad (11)$$

If we assume that the Guth chaotic inflation model [1] will be appropriate for setting a value for when the sub division of divided space into regions of $H^{-1}$ lead to classical fluctuations of the inflaton field $\phi$ being of the order of magnitude of the quantum fluctuations of the inflaton field $\phi$, we will if we use $m = .441\cdot M_P$ obtain

$$\phi^* \cong .99\cdot\pi \qquad (12)$$

{   Place  figure 1   about here  }

where we should note that equation 12 is still in the neighborhood of the $\tilde{\phi}_0$ value picked in equation 1 . This is in itself not surprising and indicates that there is some overlap in values with the simple model of inflation Guth talked about [1] . Interesting enough, this same value of the inflaton field will lead to , as seen in figure 1 , a tipping point between the true and false vacuum minimum values when we are, here, using the Bogomil'nyi inequality with

$$V(\phi_F) - V(\phi_T) \cong .041 \propto L^{-1} \tag{13}$$

which is part of

$$\frac{(\{\})}{2} \equiv \Delta E_{gap} \equiv V_E(\phi_F) - V_E(\phi_T) \tag{14}$$

and

$$\{\} \equiv \{\}_A - \{\}_B \equiv 2 \cdot \Delta E_{gap} \tag{15}$$

and

$$\{\}_A \approx \frac{m^{-2}+1}{2m^{-2}}$$

$$\{\}_B \approx \frac{\phi_T \cdot \phi_F}{6} \cdot M_P^2 \to \frac{\phi_T \cdot \phi_F}{6} \tag{16a,b}$$

We are assuming that the net topological charge will vanish and that for a 1+1 dimensional model we will be able to work with the situation as outlined in figure 2

{ Place figure 2 about here }

where the quantity in brackets is set by the developments shown in figure 1 as well as a pop up of a nucleated state as presented in the 2$^{nd}$ figure. The details of that pop up are such that we are assuming a toy model with a prototype thin wall approximation to a topological S-S' pair equivalent to assuming that the false vacuum paradigm of Sidney Coleman [2] , ( as well as Lee and Weinbergs topological solitons associated with a vacuum manifold SO(3)/ U(1) [11] ) holds in the main part.

Our contribution lies in making sense as to why one would want to have $\phi^* \cong .99 \cdot \pi$ stated as being both influential in the classical and quantum models. If one looks at figure 1, the reason for this is obvious. We should note that the value of $\phi^* \cong .99 \cdot \pi$ is at about the point where our physical system would be either tipping toward either the false or the true vacuum minimums, assuming that the bogomil'nyi inequality [5] is pertinent toward setting up minimum values for the potential in this toy model. In addition as we will argue in section III our potential model obeys the flat slow roll [4] condition even if a nucleation via a **S-S'** pair is used.

### II. Intepretation of a rate equation using this wavefunctional model of nucleation.

In the self interaction potential for the 'tunneling' scalar field, Coleman and De Luccia [3] derived a bubble nucleation rate of the form [4]

$$\Gamma_{f \to i} \equiv A \exp[-S_b + S_t] \qquad (17)$$

with

$$S_t \equiv -\frac{3}{8} \cdot \rho_t \qquad (18)$$

where one could have taken the non super-Plankian value of ( assuming $M_P \equiv 1$)

$$\rho_t \geq \frac{60}{4 \cdot \pi} \cdot M_P^2 \cdot m^2 \to \frac{60}{4 \cdot \pi} \cdot m^2 \qquad (19)$$

as well as assuming that $S_b$ was a Coleman style bounce least action integral. Here, this still though would be for a comparatively flat universe model. It is unlikely though that near the nucleation of the big bang one could escape gravitational curvature of space.

Garriga [4], assuming a nearly flat De Sitter universe also came up with an expression for the number density of particles *per unit length* ( which is time independent ) of the form

$$n \approx \frac{1}{2 \cdot \pi} \cdot \sqrt{M^2 + e \cdot \frac{E_0^2}{H^2}} \cdot \exp(-S_E) \tag{20}$$

where for our purposes we would set

$$M \leq M_P \to 1 \tag{21}$$

as well as down play the role of the applied electric field. Here, the $S_E$ is assumed to be a Euclidian action integral which would be in our example a 1+1 dimensional space knocked down to functionally being quasi 1 dimensional in 'character'. Should we assume that the per unit length is actually in reference to a Planckian length of the form

$$l_P = \left(\frac{\hbar \cdot G}{c^3}\right) \equiv 1.616 \times 10^{-35} \, meters \to 1 \tag{22}$$

where we are assuming the re scaling of $\hbar \equiv c \equiv G \equiv 1$ which we picked when we re scaled the planck time to be

$$t_P = \left(\frac{\hbar \cdot G}{c^5}\right) \to 1 \tag{23}$$

in our assumptions about the nucleation process being almost time independent and of the smallest duration possible when we discussed the formation of the wave functional used in our Appendix I discussion. This assumption effectively permitted us to reduce the de facto calculation from !+1 to being quasi one dimensional which fitted our nucleation requirements and also was in sync w.r.t. calculational convenience were we assumed a S-S' thin wall style model for the 'bubble' of space time nucleated at the beginning of creation.

Should we take into account barrier penetration directly, we can make a comparison with a net particle density from a calcuation of the form [9]

$$J \propto T_{if} \cong \frac{(\hbar^2 \equiv 1)}{2 \cdot m_e} \int \left( \Psi^*_{initial} \frac{\delta^2 \Psi_{final}}{\delta\phi(x)_2} - \Psi_{final} \frac{\delta^2 \Psi^*_{initial}}{\delta\phi(x)_2} \right) \vartheta(\phi(x)-\phi_0(x)) \wp \phi(x) \quad (24)$$

The problem we have though is in doing this we need to understand how to model variations of the phase $\phi_0$ between the $\phi_F$ and $\phi_T$ in curved space time as well as having some modifications put into the derived expression used in CDW transport [9]:

$$|T_{IF}| \approx \frac{C_1 \cdot C_2}{m^*} \cdot \left( \cosh\left(2\sqrt{\frac{x}{2L}} - \sqrt{\frac{L}{2x}}\right) \right) \cdot e^{-\alpha \cdot L \cdot \left[\frac{L}{2x}\right]} \quad (25)$$

where in this case we would have

$$m^* \equiv 2m_e \equiv 8.676 \times 10^{-20} \cdot M_P \rightarrow 8.676 \times 10^{-20} \quad (26)$$

which would make a huge difference, provided that we did not also use the normalizations of the intial and final wave functionals of the form

$$C_i = \frac{1}{\sqrt{\sqrt{\frac{L^2}{2 \cdot \pi}} \int_0^{} \exp(-2 \cdot \{\}_i \cdot \phi^2(k_N)) \cdot d\phi(k_N)}} \quad (27)$$

where the basis wave functionals would be a thin wall approximation as was done in the S-S' case done in our prior publication for CDW transport. Undoubtedly, as mentioned in the prior paper, one of the normalization constants would be quite small which would go a long way toward neutralizing the very large term due to the denominator contribution of $m^* \equiv 2m_e \equiv 8.676 \times 10^{-20} \cdot M_P \to 8.676 \times 10^{-20}$. We also would have that the length $L$ would be a de facto nucleated 'diameter' of an initially nucleated 'universe' which says that the beginning would not be a singularity as has been postulated by certain cosmologists. What we would need to work out would be at what vantage point we would set $x$ being w.r.t nucleation. The idea we are working on is that we would have in tandem with setting $L \equiv \Delta E \cong (.041)^{-1} \cong 24.39$ we are setting for the time being.

$$x = V(\phi_F) = .663 \tag{28}$$

That would not be a trivial matter to confirm. We also need to investigate the effects of curvature upon the 'evolution' phase values between the true and false vacuum states. Still though, if this were done correctly, and if we used a net Planck length as the spatial discretization of space time we were using, the results of equation 25 would not be more than an order of magnitude different from equation 20.

### III. Flat slow roll regions and the Coleman-de Luccia vacuum bubble model

We will verify existence of a region which is the flat slow roll requirement [4] of $\left|\frac{\partial^2 V}{\partial \phi^2}\right| \ll H^2$ where $H$ is the expansion rate which is a requirement of realistic inflation models [4]. This is necessary and contravenes earlier results by Coleman-de Lucia [3,4] $\delta \approx |V''|^{-1/2}$ of a thickness of a space time bubble which necessitates a phase region with $\left|\frac{\partial^2 V}{\partial \phi^2}\right| \gg H^2$. We will examine our model in terms of these two inequalities as

well as the situation as presented in figure 1 . To begin this analysis , we will definitely use Guth's [1, 3, 4] characterization of the time dependent Hubble parameter as

$$H^2 \equiv \frac{8 \cdot \pi}{3} \cdot G \cdot V(\phi) \to \frac{8 \cdot \pi}{3} \cdot V(\phi) \tag{29}$$

where we are using the convention of $G=1$ as a scaling convention . Here in our model, we may use the following. Assume we are working with $V(\phi) \to V_1(\phi)$. If so then we may look at

$$\left|\frac{\partial^2 V}{\partial \phi^2}\right|_{\phi \equiv \phi_T} = .504 << \frac{8 \cdot \pi}{3} \cdot V(\phi_T) = 4.962 \tag{30}$$

at the true vacuum position as opposed to when we have the false vacuum phase. The weird event here is that we are using the false vacuum hypothesis, but we get around the flatness problem which was so insolvable by the Coleman-de Luccia argument [3,4] . We need to note that even at the peak of the very small hill in figure 1 as well as at

$$\left|\frac{\partial^2 V}{\partial \phi^2}\right|_{\phi \equiv \phi_F} = .575 << \frac{8 \cdot \pi}{3} \cdot V(\phi_F) = 5.305 \tag{31}$$

as well as

$$\left|\frac{\partial^2 V}{\partial \phi^2}\right|_{\phi \equiv \phi^*} = .335 << \frac{8 \cdot \pi}{3} \cdot V(\phi^*) = 8.378 \tag{32}$$

for $\phi^* \equiv .99 \cdot \pi$ the slow roll condition still holds. We believe that this is in tandem with our figure 1 being somewhat simlar to the one field model of open inflation [12]

### IV. Negative pressure ? Does it still fit with our model ?

Yes it does. To see this, we should take a look at the general analysis of negative pressure we may write up as [10]

$$\in \equiv \frac{\tilde{M}_P^2}{2} \cdot \left(\frac{\partial V/\partial \phi}{V}\right)^2 << 1 \tag{33}$$

as well as

$$\eta \equiv \tilde{M}_P^2 \cdot \left(\frac{\partial^2 V/\partial \phi^2}{V}\right) << 1 \tag{34}$$

where

$$\tilde{M}_P \equiv M_P/\sqrt{8 \cdot \pi} \to 1/\sqrt{8 \cdot \pi} \tag{35}$$

we find that if $V \to V_1(\phi)$ that equations 34 and 35 definitely hold for when $\phi \to \phi_T$ as well as when $\phi \to \phi_F$ using the values of $\phi_T$ and $\phi_F$ model constraints we acquired using the Bogomil'nyi inequality showing up in the results of equations 10 and 11 . If we, instead no longer make the transformation of $V \to V_1(\phi)$, then the values of $\phi \to \phi_T$ and $\phi \to \phi_F$ as given in equations 10 and 11 no longer hold. We do make this identification explicitly $V \to V_1(\phi)$ explicit and stated so that we will be able to have equations 34 and 35 hold so that what we have is consistent with respect to known cosmological requirements.

### V.    String theory and the behavior of our scalar field $\phi$

We can refer to a basic relationship between our scalar field $\phi$ and the strength of all forces [12] gravitational and gauge alike via a relationship given by Veneziano :

$$l_P^2/\lambda_S^2 \approx \alpha_{gauge} \sim e^\phi \tag{36}$$

where the weak coupling region would correspond to where $\phi << -1$ and $\lambda_S$ is a so called quanta of length , and $l_P \equiv c \cdot t_P \sim 10^{-33} cm$. As Veneziano implies by his 2nd figure [12], a so called scalar *dilaton* field with these constraints would have behavior seen by the right hand side of figure one, with the $V(\phi) \to V(\phi_T) \approx \varepsilon^+ \geq 0$ but would have no guaranteed *false* minimum $\phi \to \phi_F < \phi_T$ and no $V(\phi_T) < V(\phi_F)$. The typical string models assume that we have a present equilibrium position in line with strong coupling corresponding to $V(\phi) \to V(\phi_T) \approx \varepsilon^+ \geq 0$ but no model corresponding for potential barrier penetration from a false vacuum state to a true vacuum in line with Colemans presentation [3]. However, if we take this to its conclusion via considerations of the FRW cosmology [13] to obtain if we start with

$$t_P \sim 10^{-42} \sec onds \Rightarrow size \ of \ universe \approx 10^{-2} cm \qquad (37)$$

which is still huge for an initial starting point, whereas we manage to in our **S-S'** 'distance model' to imply a far smaller but still non zero radii for the initial 'universe' in our toy model.

### VI. Tie in with Starboinksky model in Schrodinger description paper

In 1999, S. Biswas et al [14] wrote a very clever paper in which they managed to reduce the Wheller- De Witt equation linked to Starobinsky model to a Schrodinger equation which is time dependent. In doing this, quantum instability of de Sitter Space time [15] is essentially elaborated upon with, more importantly, a Gaussian wavefunction *anzaz* being assumed for the wave function of the '*universe*' in question. This development was justified in the paper as being necessary to accommodate what was called Hawkings style initial conditions. The tie in with our development is indirect but compelling since we derived the gaussian wavefunctional as a pre cursor to meeting initial conditions partly specified by the Bogomil'nyi inequality with respect to nucleation of an expanding universe from a very small radii , from a purely geometric-conservation law approach. Santamato [16] by way of comparison used an 'inverse' approach toward deriving similar results but from a classical Hamilton – Jacobi initial starting point.

VII.  **How dark matter ties in , using pure kinetic k essence as dark matter Template**

We shall define k essence as any scalar field with non-Cannonical kinetic terms. Following Scherrer [7], we will introduce a momentum expression via

$$p = V(\phi) \cdot F(X) \tag{38}$$

where we will define the potential in the manner we have stated for our simulation as well as set

$$X = \frac{1}{2} \cdot \nabla_\mu \phi \, \nabla^\mu \phi \tag{39}$$

and use a way to present $F$ expanded about its minimum and maximum

$$F = F_0 + F_2 \cdot (X - X_0)^2 \tag{40}$$

where we define $X_0$ via $F_X\big|_{X=X_0} = \dfrac{dF}{dX}\bigg|_{X=X_0} = 0$ as well as use a density function

$$\rho \equiv V(\phi) \cdot [2 \cdot X \cdot F_X - F] \tag{41}$$

where we find that the potential neatly cancels out of the given equation of state so

$$w \equiv \frac{p}{\rho} \equiv \frac{F}{2 \cdot X \cdot F_X - F} \tag{42}$$

as well as a growth of density perturbations terms factor Garriga and Mukhanov[17] wrote as

$$C_x^2 = \frac{(\partial p / \partial X)}{(\partial \rho / \partial X)} \equiv \frac{F_X}{F_X + 2 \cdot X \cdot F_{XX}} \qquad (43)$$

where $F_{XX} \equiv d^2 F / dX^2$, and since we are fairly close to an equilibrium value, we will pick a value of $X$ close to an extremal value.

$$X = X_0 + \varepsilon \qquad (44)$$

where if we make an averaging approximation of the value of the potential due to figure 2, as very approximately a constant, we may write the equation for the k essence field as taking the form ( where we assume $V_\phi \equiv dV(\phi)/d\phi$ )

$$(F_X + 2 \cdot X \cdot F_{XX}) \cdot \ddot{\phi} + 3 \cdot H \cdot F_X \cdot \dot{\phi} + (2 \cdot X \cdot F_X - F) \cdot \frac{V_\phi}{V} \equiv 0 \qquad (45)$$

as approximately

$$(F_X + 2 \cdot X \cdot F_{XX}) \cdot \ddot{\phi} + 3 \cdot H \cdot F_X \cdot \dot{\phi} \cong 0 \qquad (46)$$

which may be re written as

$$(F_X + 2 \cdot X \cdot F_{XX}) \cdot \ddot{X} + 3 \cdot H \cdot F_X \cdot \dot{X} \cong 0 \qquad (47)$$

which if we use equation 44 will lead to

$$\dot{\varepsilon} \cong -3 \cdot H \cdot \varepsilon \qquad (48)$$

which if we use Jaume Garrigas[17] value for $H^2 \equiv \frac{8 \cdot \pi}{3} \cdot G \cdot V(\phi) \to \frac{8 \cdot \pi}{3} \cdot V(\phi)$

will lead to

$$\varepsilon \approx \tilde{\varepsilon}_0 \cdot \exp(-8 \cdot \pi \cdot V(\phi) \cdot t) \approx \tilde{\varepsilon}_0 \cdot \exp(-8 \cdot \pi \cdot V_0 \cdot t) < \tilde{\varepsilon}_0 \qquad (49)$$

where $V_0 \approx .75 - .8$ in scaled value. In addition, we should keep in mind that we are looking at a situation where a quick nucleation will happen with $t \to t_P \sim 10^{-42} \sec onds$, which in this situation we chose to normalize as having $t \to t_P = 1$ when we picked our value for $G$ earlier on. This means that we have in this situation that we have a very small value for the 'growth of density pertubations'

$$C_S^2 \cong \frac{1}{1 + 2 \cdot (X_0 + \varepsilon) \cdot \left(1 + \frac{X_0}{\varepsilon}\right)} \approx \frac{1}{1 + 4 \cdot X_0 \left(1 + \frac{X_0}{2 \cdot \tilde{\varepsilon}_0}\right)} \quad (50)$$

if we can approximate

$$X_0 \approx \frac{1}{2} \cdot \left(\frac{\partial \phi}{\partial x}\right)^2 >> \tilde{\varepsilon}_0 \quad (51)$$

$$0 \leq C_S^2 \approx \varepsilon^+ << 1 \quad (52)$$

and [18]

$$w \equiv \frac{p}{\rho} \equiv \frac{F}{2 \cdot X \cdot F_X - F} \cong \frac{-1}{1 + 4 \cdot X_0 \cdot \left(\frac{F_2}{F_0} \cdot \tilde{\varepsilon}_0\right)} \approx -1 \quad (53)$$

This means that the initial conditions we are hypothesizing are in line with the equation of state conditions appropriate for a cosmological constant but near zero effective sound speed. This would eliminate having the initial state as behaving like pure radiation state as Casrdone et al postulated. However, should we have a situation where $X_0 \approx \frac{1}{2} \cdot \left(\frac{\partial \phi}{\partial x}\right)^2 >> \tilde{\varepsilon}_0$ no longer holds, we can have a hierarchy of evolution of the universe state as being first radiation dominated, then dark matter, and finally dark energy.

## VIII. Conclusions

We have managed to find an argument for a newly nucleated universe to have a finite but quite small diameter as well as reconcile the chaotic inflationary model of Guth[1,2] with a new fate of the false vacuum paradigm for nucleation at the initial stages of the big bang. In addition, the parameter $\phi^* \cong .99 \cdot \pi$ being where the classical and quantum effects have about the same order of magnitude in Guths inflation model[1,2] are seen in figure 1 to be the 'tipping' point between false and true vacuum values of nucleation of matter states after the big bang, where we make the assumption that the scaling we use implicitly assumes a per unit nucleation time of about the magnitude of Plancks time $t_P = \left(\dfrac{\hbar \cdot G}{c^5}\right) \to 1$ in the nucleation of matter states we assume in our least action calculation. We also speculate that if we sub divide our space time continuum to have a per unit length discretization of Plancks length $l_P = \left(\dfrac{\hbar \cdot G}{c^3}\right) \to 1$ that the value of particle creation per unit length will be of the same order of magnitude of current density as was transmitted by a re do of what was done in the CDW **S-S'** nucleation rate calculation done in the 2$^{nd}$ paper of this series[9]. We also manage to do this using a false vacuum construction which appears to be satisfying the slow roll condition w.r.t. expansion rates even if the false vacuum hypothesis is being adhered to [2,3,4]. This construction we also have initiated, has by the device of assuming an initial universe being a 'toy model' of a **S-S'** nucleation gives us a near zero but still finite 'radial' starting point which justifies our use of equations 7 and 8 above for our potential model, with the 'distance' **L** treated as a by product of the Bogomil'nyi inequality as a starting quantization condition . We also did all of this while at least maintaining at the false and true vacuum values of our scalar field potential behavior not out of sync with traditional models of negative pressure which is seen as de rigor for proper cosmological models involving inflation.

Furthermore, we also have a situation for which we can postulate an early universe which is NOT necessarily radiation dominated as Carbone et al [18] postulated, but any reading on this will await how we have modeled our initial 'kinetic' energy behavior of phase evolution in the first instance of nucleation at the value of plancks time for construction of a universe from a 'vacuum state. Further developments of our model await comprehensive evaluations of our values of $X_0 \approx \frac{1}{2} \cdot \left( \frac{\partial \phi}{\partial x} \right)^2$ at the moment of a nucleation process and may be the subject of a future paper. We need to gain more structure in how we set up suitable phase values $\phi$ which will shed light on if or not this model will indicate either a matter or radiation dominated early universe. In addition, our kinetic model can be compared with the very interesting Chimentos [19] purely kinetic k –essence model , to determine density fluctuation behavior at the initial start of a nucleation process, and to see if our density would reach $\rho =$ constant after passing through the tunneling barrier we postulate for our problem. Doing this appropriately means resolving an optimum value for the postulated $\phi$ we use in our calculations.

## Appendix I  Wave functional procedure used in our cosmology example

Traditional current treatments frequently follow the Fermi golden rule for current density

$$J \propto W_{LR} = \frac{2 \cdot \pi}{\hbar} \cdot |T_{LR}|^2 \cdot \rho_R(E_R) \tag{1}$$

In our prior work we applied the Bogomil'nyi inequality [5,9] to come up with an acceptable wave functional which will refine *I-E* curves used in density wave transport. We shall, here generalize what was in the Guth paper [7] a de facto 1+1 dimensional problem in cosmology to being one which is quasi one dimensional by making the following substitution, namely looking at the lagrangian density $\varsigma$ to having a time independent behavior denoted by a sudden pop up of a S-S' pair via the substitution of the nucleation 'pop up' time by

$$\int d\tau \cdot dx \cdot \varsigma \to t_P \cdot \int dx \cdot L \tag{2}$$

where $t_P$ is here the Planck's time interval. Then afterwards, we shall use the substitution of $\hbar \equiv c \equiv 1$ so we can write

$$\int d\tau \cdot dx \cdot \varsigma \to t_P \cdot \int dx \cdot L \equiv G \cdot \int dx \cdot L \tag{3}$$

where

$$M_P \equiv \frac{1}{\sqrt{G}} \equiv 1.22 \times 10^{19} \, GeV = .231 \times 10^{20} \cdot m_e \quad (4)$$

such that

$$m_e \equiv 4.338 \times 10^{-20} \cdot M_P \to 4.338 \times 10^{-20} \quad (5)$$

So, if we make the substitution that $M_P \equiv 1 \Rightarrow G \equiv 1$ as a normalization procedure, we have

$$\int d\tau \cdot dx \cdot \varsigma \to t_P \cdot \int dx \cdot L \equiv \int dx \cdot L \quad (6)$$

This allowed us to use in our cosmology nucleation problem the following wave functional

$$\psi \propto c \cdot \exp\left(-\beta \cdot \int L \, dx\right) \quad (7)$$

in a functional current we derived as being of the form

$$J \propto T_{if} \quad (8)$$

when

$$T_{if} \cong \frac{(\hbar^2 \equiv 1)}{2 \cdot m_e} \int \left( \Psi_{initial}^* \frac{\delta^2 \Psi_{final}}{\delta \phi(x)_2} - \Psi_{final} \frac{\delta^2 \Psi_{initial}^*}{\delta \phi(x)_2} \right) \vartheta(\phi(x) - \phi_0(x)) \wp \phi(x) \quad (9)$$

with the electron mass re written via the conventions of equation 5 and where

$$c_2 \cdot \exp\left(-\alpha_2 \cdot \int d\tilde{x} [\phi_T]^2\right) \cong \Psi_{final} \quad (10)$$

and

$$c_1 \cdot \exp\left(-\alpha_1 \cdot \int dx [\phi_0 - \phi_F]^2\right) \equiv \Psi_{initial} \quad (11)$$

with $\phi_0 \equiv \phi_F + \varepsilon^+$ and where the $\alpha_2 \cong \alpha_1$ . These values for the wave functionals showed up in the upper right hand side of figure 1 (as well as figure 2 )and represent the decay of the false vacuum hypothesis which we found was in tandem with the Bogomil'nyi inequality[5,9]. As mentioned in our prior papers [6,9] this allows us to present a change in energy levels to be inversely proportional to the distance between a S-S' pair

$$\alpha_2 \equiv \Delta E_{gap} \equiv \alpha \approx L^{-1} \tag{12}$$

We also found that in order to have a gaussian potential in our wavefunctionals that we needed to have

$$\frac{(\{\ \})}{2} \equiv \Delta E_{gap} \equiv V_E(\phi_F) - V_E(\phi_T) \tag{13}$$

where for potentials of the form ( generalization of the extended sine Gordon model potential )

$$V_E \cong C_1 \cdot (\phi - \phi_0)^2 - 4 \cdot C_2 \cdot \phi \cdot \phi_0 \cdot (\phi - \phi_0)^2 + C_2 \cdot (\phi^2 - \phi_0^2)^2 \tag{14}$$

as a template for analyzing

$$V_1(\phi) = M_P^2 \cdot (.5989) \cdot (1 - \cos(\phi)) + \frac{m^2}{2} \cdot (\phi - \phi^*)^2 \tag{15}$$

when we are looking at :

$$V(\phi) \equiv \left[ initial\ energy\ density \right] + V_1(\phi) \tag{16}$$

where we normalize out the initial energy density in our treatment of the wave functionals. But in this procedure we had a lagrangian [6,9] we modified to be ( due to the Bogomil'nyi inequality )

$$L_E \geq |Q| + \frac{1}{2} \cdot (\phi_0 - \phi_C)^2 \cdot \{ \} \qquad (17)$$

with topological charge $|Q| \to 0$ and with the gaussian coefficient found in such a manner as to leave us with wave functionals [6,9] we generalized for charge density transport

$$\Psi_f[\phi(\mathbf{x})]\big|_{\phi \equiv \phi_{Cf}} = c_f \cdot \exp\left\{ -\int d\mathbf{x}\, \alpha \left[ \phi_{Cf}(\mathbf{x}) - \phi_0(\mathbf{x}) \right]^2 \right\}, \qquad (18)$$

and

$$\Psi_i[\phi(\mathbf{x})]\big|_{\phi \equiv \phi_{Ci}} = c_i \cdot \exp\left\{ -\alpha \int d\mathbf{x} \left[ \phi_{ci}(\mathbf{x}) - \phi_0 \right]^2 \right\}, \qquad (19)$$

**Appendix II.  Basis function assuming a thin wall approximation to a S-S' pair**

In momentum space, the following 'thin wall' approximation [9] was used for our example

$$\phi(k_n) = \sqrt{\frac{2}{\pi}} \cdot \frac{\sin(k_n L/2)}{k_n} \qquad (1)$$

is the F.T. of a 'box. of length $L$ and of height $2 \cdot \pi$ which is a drastically re scaled model of initial matter states at the beginning of nucleation of a new universe. This assumes that the initial states of matter had zero initial charge . We used this as well with different length assumptions for our S-S' pair production in our CDW transport problem.

# Figure captions

**Fig 1 :** Evolution from an initial state $\Psi_i[\phi]$ to a final state $\Psi_f[\phi]$ for a tilted double-well potential in a quasi 1-D cosmological model for inflation, showing a kink-antikink pair bounding the nucleated bubble of true vacuum. This illustrates the direct influence of the Bogomil'nyi inequality in giving a linkage between the 'distance' between constituents of a cosmological 'nucleated pair' of **S-S'** and the $\Delta E$ difference in energy values between $V(\phi_F)$ and $V(\phi_T)$ which allowed us to have a 'gaussian' representation of evolving nucleated states.

**Fig 2 :** Evolution from an initial state $\Psi_i[\phi]$ to a final state $\Psi_f[\phi]$ for a double-well potential (inset) in a quasi 1-D model, showing a kink-antikink pair bounding the nucleated bubble of true vacuum. The shading illustrates quantum fluctuations about the optimum configurations of the field $\phi_F$ and $\phi_T$, while $\phi_0(x)$ represents an intermediate field configuration inside the tunnel barrier

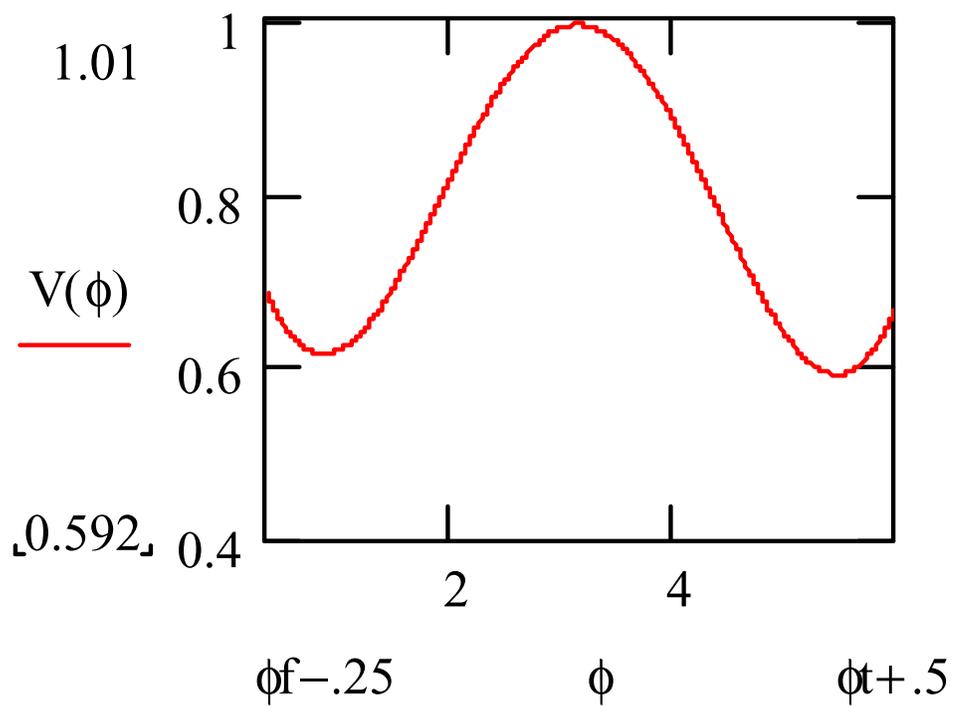

Figure 1

Beckwith et al.

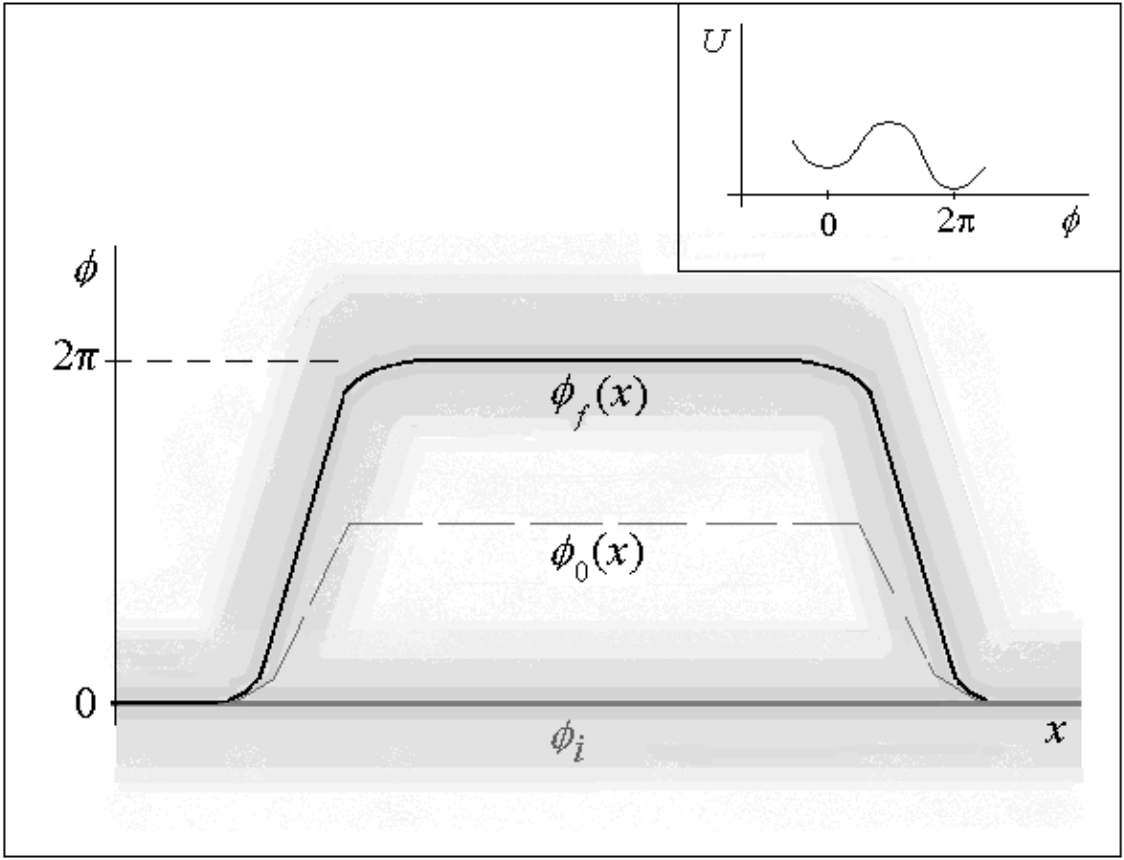

Figure 2

Beckwith et al